\documentclass[aps,pra,twocolumn,groupedaddress,showpacs]{revtex4}
\usepackage{graphicx}
\begin{document}
%
\title{Mixed state sensitivity of several quantum information benchmarks}
\author{Nicholas A. Peters} 
\author{Tzu-Chieh Wei}
\author{Paul G. Kwiat}
\affiliation{Physics Department, University of Illinois, 1110 West Green Street, Urbana, IL 61801}
\date{July 21, 2004}
\begin{abstract}
We investigate an imbalance between the sensitivity of the common state measures--fidelity, trace distance, concurrence, tangle, von Neumann entropy and linear entropy--when acted on by a depolarizing channel.  Further, in this context we explore two classes of two-qubit entangled mixed states.  Specifically, we illustrate a sensitivity imbalance between three of these measures for depolarized (i.e., Werner-state like) nonmaximally entangled and maximally entangled mixed states, noting that the size of the imbalance depends on the state's tangle and linear entropy.   
\end{abstract}

\maketitle
Because the outcome of most quantum information protocols hinges on the quality of the initial state, pure maximally entangled states are often the optimal inputs.  However, decoherence and dissipation inevitably decrease the purity and entanglement of resource states, yielding partially entangled mixed states.  The most common measure used to benchmark a starting state resource is the fidelity~\cite{jozsa94}, as used e.g., in entanglement purification~\cite{bennett96b, kent98} and optimal mixed state teleportation~\cite{vv03}.  Likewise, the success of these procedures is often judged using the fidelity of the output state with some target, as is the case, for example, in quantum cloning~\cite{gisin97}.  Recently it was found that, for the specific case of maximally entangled mixed states~\cite{ih01,vad01,munro01}~(MEMS), using the fidelity to compare an experimentally produced state and a target state was a less sensitive way of assessing experimental agreement than comparing the tangle~\cite{wooters98,coffman00} and the linear entropies~\cite{bose00} of those states~\cite{peters04}.  Because one needs to understand the best way to benchmark states for quantum information protocols, here we examine the fidelity for more general entangled two-qubit mixed quantum states and note its behavior in relation to the common state measures of linear and von Neumann entropy, tangle and concurrence, and trace distance.  

After some general calculations for depolarized states, we consider explicitly two classes of two-qubit entangled states acted on with depolarizing channels: nonmaximally entangled states and maximally entangled mixed states.  The effect of a depolarizing channel is to make the states we study similar to the Werner states (an incoherent combination of a pure maximally entangled state and completely mixed state)~\cite{werner,footnotewerner}, which have been realized with polarized photons~\cite{zhang02,barbieri04}.  These two classes of states were chosen because they allow us to study mixed state entanglement for states of current interest, and also to understand how these states change under uniform depolarization.  Such a uniform depolarization model is applicable to many examples of real experimental decoherence.

\section{General sensitivities of measures}
Before considering specific examples of entangled mixed states, we examine general sensitivities for several measures using generic depolarized density operators.
The depolarized $N$-level system ($N=2$ for a qubit, $N=4$
for two qubits, etc.) is
\begin{equation}
\rho\rightarrow \rho'=(1-\epsilon)\rho + \frac{\epsilon}{N}\openone_N,
\label{petrho}
\end{equation}
where $\epsilon$ is the strength of depolarization.

\subsection{Fidelity}
For direct comparison of two mixed states, e.g., $\rho_t$ and $\rho_p$, for target and perturbed states, respectively, we first discuss the fidelity introduced by Jozsa~\cite{jozsa94}: 
\begin{equation}
F(\rho_t,\rho_p)\equiv \left| {\rm
Tr}\left(\sqrt{\sqrt{\rho_t}\rho_p\sqrt{\rho_t}}\right) \right|^2.
\label{fidelity}
\end{equation}
In the simpler case of two pure states $|\psi_t\rangle$ and $|\psi_p\rangle$, $F$ reduces to $|\langle \psi_p |\psi_t \rangle|^2$.  It is also important to note that some researchers, as in~\cite{nc00}, use an amplitude version of the fidelity:  $f\equiv\sqrt{F}$.  In either case, the fidelity is zero for orthogonal states and one for identical states.

Because we wish to consider small perturbations in the fidelity, the ``amplitude version'' $f$ should be less sensitive because it lacks the square.  We consider a generic state $\rho$ with eigenvalues $\{\lambda_i\}$, depolarized by $\epsilon$.
The amplitude fidelity $f$ between the output state $\rho'$ and the
input $\rho$ is
\begin{eqnarray}
f(\rho,\rho')&=&{\rm Tr}\sqrt{(1-\epsilon)\rho^2+\frac{\epsilon}{N}\rho}\\
&=&\sum_{i}\sqrt{(1-\epsilon)\lambda_i^2+\frac{\epsilon}{N}\lambda_i}.
\end{eqnarray}
We assume $\epsilon$ is small such that $\epsilon \ll N\lambda/|1-N\lambda|$,
where $\lambda$ is the smallest nonzero eigenvalue.  Thus, we can
expand the above expression to second order in $\epsilon$:
\begin{eqnarray}
&&f\approx\sum_{\lambda_i\ne0}\lambda_i\left\{ 1+\frac{1}{2}\left( \frac{1\!-\!N\lambda_i}{N\lambda_i}  \right)\epsilon
-\frac{1}{8}\left(\frac{1\!-\!N\lambda_i}{N\lambda_i}\right)^2\epsilon^2\right\} \nonumber\\
\!\!\!\!\!\!\!\!\!\!&&=1-\left(\frac{1}{2}-\frac{n_\otimes}{2N}\right)\epsilon-\sum_{\lambda_i\ne0}\frac{\lambda_i}{8}\left(\frac{1\!-\!N\lambda_i}{N\lambda_i}\right)^2\epsilon^2+{\cal O}(\epsilon^3),\nonumber\\&&
\label{Fexpand}
\end{eqnarray}
where $n_\otimes$ is the number of nonzero eigenvalues of $\rho$.
When $\rho$ is of full rank (i.e., $n_\otimes=N$), the first order
term vanishes, and the fidelity is sensitive only to second order
in the small depolarizing parameter. If $\rho$ is not full rank, 
$f$ {\em is} sensitive to first order, but becomes less so as the rank becomes higher.
Squaring the result~(\ref{Fexpand}) in fact gives the {\em same} order of sensitivity for $F$.

\subsection{Trace distance}
Another possible measure used to compare two states is the trace distance~\cite{nc00}, given by
\begin{equation}
D(\rho_t,\rho_p)\equiv\frac{1}{2}{\rm Tr}|\rho_t-\rho_p|.
\label{trace_distance}
\end{equation}
Evaluating the trace distance using (\ref{petrho}) gives
\begin{eqnarray}
D(\rho,\rho')=\frac{1}{2}\sum_i \mid\lambda_i-\frac{1}{N}\mid\epsilon.
\end{eqnarray}

Here the $1/N$ term comes from the $N\times N$ mixed state ($\openone_N/N$) used to depolarize $\rho$ to create $\rho'$ (\ref{petrho}).  Thus, we see that the trace distance is always {\em linearly} sensitive
to the strength of depolarization, except for $\rho=\openone_N/N$, i.e., the fully mixed state.  Consequently, the difference between two similar states will in general be less apparent when using $f$ (or $F$) than when using $D$.

\subsection{Linear entropy}
To quantify the mixedness of a given state $\rho$, we first consider the linear entropy ($S_L$), which is based on the purity, and for an $N$-level system is
\begin{equation}
S_{\rm L}(\rho)\equiv\frac{N}{N-1}[1-{\rm Tr}(\rho^2)].
\label{generalline}
\end{equation}
The linear entropy is zero for pure states and one for completely mixed states, i.e., $S_L=1$ for the normalized $N$-qubit identity $\openone_N/N$.
The change in the linear entropy under a depolarizing channel is:
\begin{eqnarray}
\Delta S_{\rm L}\equiv S_{\rm L}(\rho')-S_{\rm L}(\rho)=(2\epsilon-\epsilon^2)(1-S_{\rm L}). 
\end{eqnarray}
Therefore, the linear entropy is always linearly sensitive
in $\epsilon$, except when $S_{\rm L}(\rho)=1$, namely, when $\rho$ is the fully mixed state $\openone_N/N$. 
Thus, the linear entropy is, in general, more sensitive to the depolarizing
channel than the fidelity, as was previously shown for the specific case of any depolarized linear pure {\em single}-qubit state~\cite{pwk04}.

\subsection{Von Neumann entropy}
Another frequently encountered entropy measure
is the von Neumann entropy:
\begin{equation}
S_{\rm V}(\rho)\equiv-{\rm Tr}\left(\rho \ln \rho\right).
\end{equation}
Using~(\ref{petrho}) and evaluating $\Delta S\equiv S_{\rm V}(\rho')-S_{\rm V}(\rho)$ 
to first order gives
\begin{eqnarray}
\Delta S&\approx&-\frac{n_0}{N}\epsilon\,\ln\,\epsilon+ \\
&&\epsilon\left(1-S_{\rm V}(\rho)-\frac{n_\otimes}{N}+\frac{n_0}{N}\ln N-  \frac{1}{N}\sum_{\lambda_i\ne0}\ln\lambda_i\right)\nonumber,
\end{eqnarray}
where $n_\otimes$ ($n_0$) is the number of nonzero (zero) eigenvalues of $\rho$,
and $n_\otimes+n_0=N$.
When $\rho$ is not a full rank matrix (i.e., $n_0\ne0$),
the von Neumann entropy is, to leading order, sensitive in $\epsilon\ln\epsilon$ (stronger
than order $\epsilon$). As the rank become higher, this $\epsilon\ln\epsilon$
sensitivity decreases. When $\rho$ is of full rank (i.e., $n_0=0$ and $n_{\otimes}=N$), the von Neumann
entropy is linearly sensitive in $\epsilon$ unless $S_{\rm V}=-\frac{1}{N}\sum_i\ln\lambda_i$, which is again possible only when $\lambda_i=1/N$, i.e.,
for the fully mixed state $\rho=\openone_N/N$.

\subsection{Concurrence and Tangle}
Here we examine two ways of quantifying the entanglement of a system, restricting our attention to two-qubit states.  We will first derive the variation of the concurrence for an entangled state acted on by a depolarizing channel, then use this to find the result for the tangle, which is the concurrence squared.

\subsubsection{Concurrence}
The concurrence is given by~\cite{wooters98}
\begin{equation}
C(\rho)\equiv max\{0,\sqrt{\lambda_1}-\sqrt{\lambda_2}-\sqrt{\lambda_3}-\sqrt{\lambda_4}\},
\end{equation}
where $\lambda_i$ are the eigenvalues of $\rho\tilde{\rho}$ in non-increasing order by magnitude.  Here we define
$\tilde{\rho}\equiv(\sigma_2 \otimes \sigma_2)\rho^{\ast}(\sigma_2 \otimes  \sigma_2)$ with 
$\sigma_2= \left( \begin{array}{cc} 0 & -i \\ i & 0 \end{array} \right)$. 

Suppose \{$\lambda_{i}$\} are arranged in
non-increasing order, and the state $\rho$ is entangled, so that
$C(\rho)=\sqrt{\lambda_1}-\sqrt{\lambda_2}-\sqrt{\lambda_3}-\sqrt{\lambda_4}$.
(If $\rho$ is unentangled, $\rho'$, which has additional noise, is still unentangled.)
To find the concurrence of $\rho'$, we have to evaluate the eigenvalues of the matrix
\begin{equation}
\rho'\tilde{\rho'}=(1-\epsilon)^2\rho\tilde\rho+\frac{\epsilon}{4}(1-\epsilon)(\rho+\tilde\rho)+\frac{\epsilon^2}{16}\openone_4.
\end{equation}
We can treat the last two terms as perturbations and evaluate the
eigenvalues to leading order:
\begin{equation}
\lambda_i'\approx(1-\epsilon)^2\lambda_i+\frac{\epsilon}{4}(1-\epsilon)
\langle\rho+\tilde\rho\rangle_i+\frac{\epsilon^2}{16},
\end{equation}
where \begin{equation}
\langle\rho+\tilde\rho\rangle_i\equiv\langle\lambda_i|(\rho+\tilde\rho)|\lambda_i\rangle.
\end{equation}
For $\epsilon<\lambda$, where $\lambda$ is the smallest nonzero value of
$\{\lambda_i\}$, we have, to leading order,
\begin{equation}
\sqrt{\lambda_i'}\approx (1-\epsilon)\sqrt{\lambda_i}+\frac{\epsilon}{8\sqrt{\lambda_i}}\langle\rho+\tilde\rho\rangle_i.
\end{equation}
Hence, the change in concurrence ($\Delta C\equiv C(\rho')-C(\rho)$), is given by
\begin{eqnarray}
\Delta C&\approx&-\sum_{\lambda_i=0}\sqrt{\frac{\epsilon}{4}(1-\epsilon)\langle\rho+\tilde{\rho}\rangle_i+\frac{\epsilon^2}{16}} -\epsilon\, C(\rho) \nonumber \\
&& +\frac{\epsilon}{8}
\left(\frac{\langle\rho+\tilde\rho\rangle_1}{\sqrt{\lambda_1}}
-\sum_{i=2,\lambda_i\ne0}^4\frac{\langle\rho+\tilde\rho\rangle_i}{\sqrt{\lambda_i}}\right).
\end{eqnarray}
The variation of concurrence is thus at worst first order in $\epsilon$ except for the unlikely case that
\begin{eqnarray}
C(\rho)=\frac{\langle\rho+\tilde\rho\rangle_1}{8\sqrt{\lambda_1}} - \sum_{i=2,\lambda_i\ne0}^4\frac{\langle\rho+\tilde\rho\rangle_i}{8\sqrt{\lambda_i}},
\end{eqnarray}
when $\rho$ is of full rank.

\subsubsection{Tangle}
To characterize a state's entanglement, one may also use the tangle~\cite{wooters98,coffman00}, i.e., the concurrence squared: 
\begin{equation}
T(\rho)~=~C(\rho)^2.
\end{equation}
Using the result for variation in concurrence, the variation of tangle can now be expressed as
$T'-T\approx 2 C\Delta C$.  Thus, the tangle is also typically sensitive in the first order to depolarizing perturbations.  

In summary, we have thus far shown that, under the influence of a small depolarizing channel, the fidelity is not as sensitive as the change in trace distance, linear entropy, von Neumann entropy, concurrence, and tangle.  Next we shall illustrate this fact for specific states and investigate the situation for larger depolarization and for variable entanglement. 

\section{Investigation for specific states}
 
The first state we consider is similar to the classic Werner state, but we allow {\em arbitrary} entanglement through the use of a variable nonmaximally entangled pure state component in addition to the mixed state dilution:
\begin{equation}
\rho_1(\epsilon, \theta)\equiv(1-\epsilon)|\Psi(\theta)\rangle \langle \Psi(\theta)|+ \frac{\epsilon}{4}\openone_4,~~{\rm with}
\label{rho1}
\end{equation}
\begin{equation}
|\Psi{(\theta)}\rangle\equiv\cos{2\theta}|00\rangle+\sin{2\theta}|11\rangle, 
\label{entangledeq}
\end{equation}
where the parameter $\theta$ controls the entanglement and $\epsilon$ the mixedness.  We choose this parameterization for simplicity and because the entropy and the entanglement of the state are somewhat uncoupled from each other.  In this case, the concurrence is $C(\rho_1(\epsilon, \theta))=\max\{0, 2(1-\epsilon)\cos 2\theta\sin 2\theta-\epsilon/2\}$ (assuming $\cos2\theta\sin2\theta\geq0$) and the linear entropy depends only on epsilon, $S_L(\rho_1(\epsilon, \theta))=2\epsilon-\epsilon^2$.  In a similar way, we depolarize a maximally entangled mixed state (MEMS) according to  
\begin{equation}
\rho_2(\epsilon, r)\equiv(1-\epsilon)\rho_{MEMS}(r)+ \frac{\epsilon}{4}\openone_4,
\label{rho2}
\end{equation}
where the MEMS, using the parameterizations of concurrence (or equivalently tangle) and linear entropy, is given by~\cite{munro01}
\begin{eqnarray}
\rho_{MEMS~I}&= \bordermatrix{ 
& \cr
&\frac{r}{2} &0 &0 &\frac{r}{2} \cr
&0 &1-r &0 &0 \cr
&0 &0 &0 &0 \cr
&\frac{r}{2} &0 &0 &\frac{r}{2} \cr}
 ,&\frac{2}{3} \leq r \leq 1,
\label{memsI} \nonumber 
\\
\rho_{MEMS~II}&= \bordermatrix{ 
& \cr
&\frac{1}{3} &0 &0 &\frac{r}{2} \cr
&0 &\frac{1}{3} &0 &0 \cr
&0 &0 &0 &0 \cr
&\frac{r}{2} &0 &0 &\frac{1}{3} \cr}
 ,&0 \leq r \leq \frac{2}{3};
\label{memsII}
\nonumber 
\end{eqnarray}
the parameter $r$ is the concurrence of the MEMS.  
\begin{figure}
\begin{center}
\includegraphics[width=8.6cm]{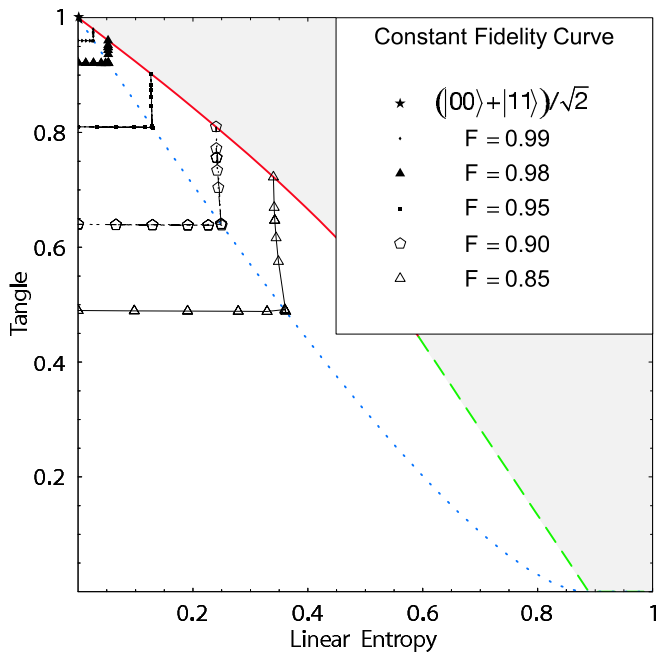}
\caption{Constant fidelity curves for the maximally entangled state $(|00\rangle+|11\rangle)/\sqrt{2}$ (star, upper left corner). Also shown are the Werner state curve (dotted line) and, bounding the gray region of nonphysical entropy-tangle combinations, the MEMS curve, which is solid for $\rho_{MEMS~I}$ and dashed for $\rho_{MEMS~II}$.  The (horizontal) constant fidelity curves below the Werner state curve are swept out by comparing the starting state with states of the form $\rho_1(\epsilon, \theta)$,  equation (\ref{rho1}), while the (nearly vertical) curves above the Werner state line are generated by varying the parameters of $\rho_2(\epsilon, r)$ given by equation (\ref{rho2}).  For comparison, the pure product state $|00\rangle$ (lower left corner) has fidelity of 0.5 with this target.}
\label{bellFcalc}
\end{center}
\end{figure}

With these parameterizations, we may map out constant fidelity curves between a target state and a perturbed state in the linear entropy-tangle plane (we choose these particular measures for calculational simplicity and because ($\ref{rho1}$) and ($\ref{rho2}$) cover the entire physically allowed region of the plane).  It is our purpose to use these curves to gain insight as to how the entanglement and mixedness may vary over a constant fidelity curve and how this variation may in turn depend on the amount of entanglement and mixedness.  To do this, we calculate the fidelity between a target state $\rho_1(\epsilon_t, \theta_t)$ and a perturbed state $\rho_1(\epsilon_p, \theta_p)$.  Specifically, the parameters $\epsilon_p$ and $\theta_p$ are varied to create perturbed states of all possible tangle and entropy values as long as the perturbed state has a given fidelity with the target.  Likewise, the process is repeated for $\rho_2(\epsilon_t, r_t)$, but instead varying the parameters of $\rho_2(\epsilon_p, r_p)$.
\begin{figure}
\begin{center}
\includegraphics[width=8.6cm]{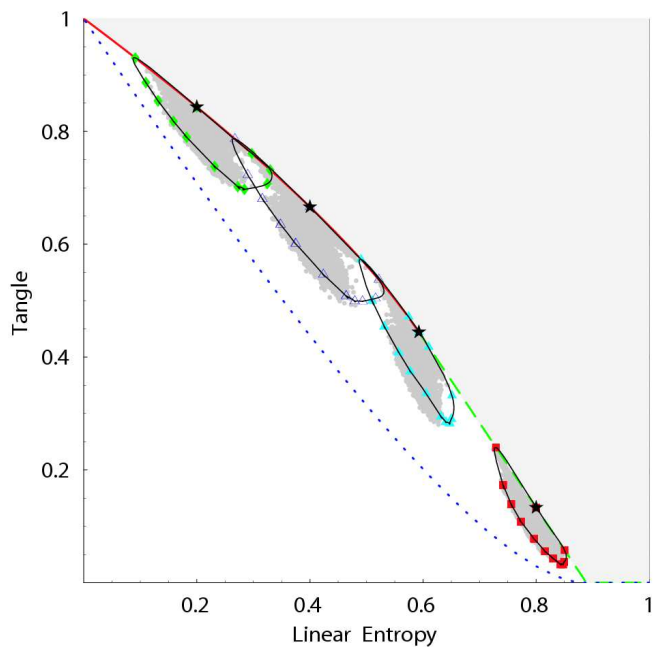}
\caption{Constant 0.99-fidelity curves for several starting MEMS, plotted as stars.  The constant fidelity curves are calculated by comparing the target state with $\rho_2(\epsilon, r)$ where $\epsilon$ and $r$ are varied to give different tangles and linear entropies.  The dark gray regions are 20000 points per initial MEMS, corresponding to numerically generated density matrices that have fidelity of 0.9900 or higher with the target.}
\vspace{-0.0cm}
\label{memsFcalc}
\end{center}
\end{figure}

In the pure, maximally entangled limit, both (\ref{rho1}) and (\ref{rho2}) reduce to the maximally entangled state $|\phi^+\rangle\equiv(|00\rangle+|11\rangle)/\sqrt{2}$.  Therefore, this is a natural state with which to start our discussion.  Because (\ref{rho1}) and (\ref{rho2}) occupy different regions of the entropy-tangle plane, it is not surprising that we need to use both equations to map out the constant fidelity curves for $|\phi^+\rangle$, as shown in Fig.~\ref{bellFcalc}.  The horizontal curves in the region bounded above by the Werner state curve, are traced out by computing the fidelity of $|\phi^+\rangle$ with (\ref{rho1}).  This fidelity is $F=(1+\sqrt{T})/2$ and, surprisingly, does not explicitly depend on the depolarization of the perturbed state.  The maximal fidelity of any two-qubit
state with maximally entangled pure states was found by Verstraete and Verschelde~\cite{fv02} to be
bounded above by $(1+\sqrt{T})/2$. The two-qubit states (\ref{rho1}) saturate this bound (as does any two-qubit pure state). Any entangled state that saturates this bound apparently has $F>1/2$, thus allowing concentration of entanglement via the BBPSSW scheme~\cite{bennett96b} without requiring local filtering~\cite{hhh97}.  Another consequence of this simple fidelity expression is that, when comparing (\ref{rho1}) with $|\phi^+\rangle$, the fidelity by itself cannot distinguish between pure nonmaximally entangled states and Werner states of the same tangle.  For example, both the nonmaximally entangled pure state $\rho_1(0, 11.25^\circ)$ and the Werner state $\rho_1(0.19525, 22.5^\circ)$ have tangle equal to 0.5, and each has fidelity 0.854 with $|\phi^+\rangle$.  

To trace the curves above the Werner state line, we calculate the fidelity of $|\phi^+\rangle$ with equation (\ref{rho2}).  In this case, the analytic expression for the fidelity does not provide much insight, so we only present numerical results, yielding the nearly vertical curves shown in Fig.~\ref{bellFcalc}.  Notice that the vertical curves scale nearly the same as the horizontal curves.  Thus, when comparing $|\phi^+\rangle$ with states created using (\ref{rho1}) and (\ref{rho2}) that each separately have the same fidelity with $|\phi^+\rangle$, the linear entropy and tangle for (\ref{rho1}) and (\ref{rho2}) change by about the same amount when the fidelity changed.  So both (\ref{rho1}) and (\ref{rho2}) display approximately the same fidelity insensitivity.

Next we consider the effect of depolarizing target maximally entangled mixed states (MEMS).  In this case, we calculate the 0.99-fidelity curve for several target states, shown as stars in Fig.~\ref{memsFcalc}.  Note that the 0.99-fidelity curve encloses a much larger area for any of the MEMS targets than it does for the $|\phi^+\rangle$ calculation (Fig.~\ref{bellFcalc}).  We attribute this to the fact that depolarizing a pure state changes the fundamental character (as measured with the fidelity) of the state more than does depolarizing an {\em already} mixed state.  Also shown in Fig.~\ref{memsFcalc} are the results of a numerical Monte Carlo simulation, where we assumed an ideal starting state, then calculated the predicted counts one would expect to measure in an experiment if there were no measurement noise or fluctuations.   These ideal counts are then perturbed in a statistical way to give a variation one might expect in an experimental measurement for a total collection of $\sim$2000 counts~\cite{footnotesimulation}.  Note that the sizes and shapes of the simulation and the constant fidelity curves are similar but not identical.  As the simulation is random, it behaves somewhat like a depolarizing channel, adding uniform noise (explaining some of the similarity); however, random fluctuations are not enough to mimic the extreme changes along the MEMS curve, as the MEMS density matrices posses a very specific form.  

\begin{figure}
\begin{center}
\includegraphics[width=8.6cm]{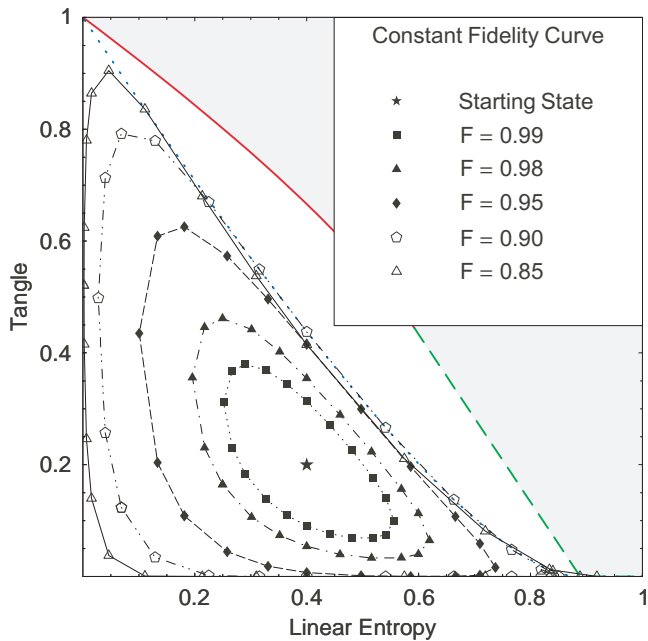} 
\caption{Constant fidelity curves for a nonmaximally entangled mixed state ($\rho_2(\epsilon=0.225,\theta=11.57^\circ$)) compared with states calculated by varying $\epsilon$ and $\theta$.  In the case of this mixed entangled state, the constant fidelity regions are surprisingly large: for example, a 0.9 fidelity with this starting state could arise from a nearly pure entangled state or from an unentangled near fully mixed state.  The 0.9 and higher fidelity region covers $\sim$60$\%$ of the physically allowed region of the linear entropy-tangle plane.}
\label{mixedFcalc}
\end{center}
\end{figure}


The two previous cases dealt with states that have the highest entanglement values, i.e., they are bounded by the edges of the physically allowed regions of the entropy-tangle plane.  To investigate the behavior ``on the open plane,'' we examine an entangled mixed target state that is a specific example of (\ref{rho1}):
\begin{eqnarray}
\rho_{1}(\sim0.225,\sim11.57^\circ)=\cr \bordermatrix{ 
& \cr
&0.7113 &0 &0 &0.2800 \cr
&0 &0.0564 &0 &0 \cr
&0 &0 &0.0564 &0 \cr
&0.2800 &0 &0 &0.1760 \cr},
\label{expampleMS}
\end{eqnarray}
which is shown as a star in Fig.~\ref{mixedFcalc}.  Note that the 0.99-fidelity region is much larger than for any of the previous target states, including the MEMS.  This result is particularly astonishing when viewed in light of what is typically considered ``high fidelity'' experimentally for entangled states: 0.9 to 0.99 depending on the particular two-qubit implementation (although some {\em single} qubit fidelities have been reported at the 0.999 level~\cite{peters03}).  Consider the 0.9-fidelity curve in Fig.~\ref{mixedFcalc}.  This level of fidelity with the target states could mean one has a nearly pure maximally entangled state ($S_L\cong 0.07, T\cong 0.79$) or a nearly completely mixed unentangled state ($S_L\cong 0.85,T=0$).  

The extreme insensitivity of the fidelity for the target state (\ref{expampleMS}) is consistent with equation~(\ref{Fexpand}), which indicates that fidelity sensitivity drops off as the rank of a state increases.  In this case (\ref{expampleMS}) has rank four while the MEMS have either rank two or three (and the maximally entangled pure state $|\phi^+\rangle$ is rank 1); thus (\ref{expampleMS}) exhibits larger constant fidelity curves than the MEMS or $|\phi^+\rangle$.  In addition, we conjecture that this effect may be further exacerbated because the addition of symmetric noise to an already highly mixed state (which has a symmetric form) changes the character of the state less than for a MEMS (which has an asymmetric form).    

In summary, we have shown an imbalance between the sensitivities of the common state measures--fidelity, trace distance, concurrence, tangle, linear entropy and von Neumann entropy--for two classes of two-qubit entangled mixed states.  This imbalance is surprising in light of the fact that orthogonal states which have zero fidelity with one another may have the same entanglement and mixedness; thus, one might have expected the fidelity to be a {\em more} sensitive means to characterize a state than quantifying state properties like entanglement and mixedness.  Here we have shown an opposite effect.  Specifically, we have investigated several examples at different locations in the entropy-tangle plane, where the trend shows progressively larger 0.99-fidelity regions as the state becomes more mixed and less entangled.  We also have shown that, at least for maximally entangled target states, the fidelity is insensitive when comparing between Werner states and nonmaximally entangled states of the same tangle.  This work has important ramifications for benchmarking the performance of quantum information processing systems, as it reveals that the usually quoted measure of fidelity is often a remarkably poor indicator, e.g., of the entanglement in a state, on which the performance of quantum information systems often depend.  This may have consequences, for example, for determining the limits of fault tolerant quantum computation~\cite{klz98}, and it may be beneficial to include other benchmarks in addition to/instead of fidelity when characterizing resources needed for various quantum information protocols.    

\begin{acknowledgments}     
The authors would like to thank Mike Goggin for comments useful in refining this manuscript.  This work was supported by the National Science Foundation (Grant number EIA-0121568), and the MURI Center for Photonic Quantum Information Systems (ARO/ARDA program DAAD19-03-1-0199).
\end{acknowledgments}
\bibliography{noise}

\begin{thebibliography}{23}
\expandafter\ifx\csname natexlab\endcsname\relax\def\natexlab#1{#1}\fi
\expandafter\ifx\csname bibnamefont\endcsname\relax
  \def\bibnamefont#1{#1}\fi
\expandafter\ifx\csname bibfnamefont\endcsname\relax
  \def\bibfnamefont#1{#1}\fi
\expandafter\ifx\csname citenamefont\endcsname\relax
  \def\citenamefont#1{#1}\fi
\expandafter\ifx\csname url\endcsname\relax
  \def\url#1{\texttt{#1}}\fi
\expandafter\ifx\csname urlprefix\endcsname\relax\def\urlprefix{URL }\fi
\providecommand{\bibinfo}[2]{#2}
\providecommand{\eprint}[2][]{\url{#2}}

\bibitem[{\citenamefont{Jozsa}(1994)}]{jozsa94}
\bibinfo{author}{\bibfnamefont{R.}~\bibnamefont{Jozsa}},
  \bibinfo{journal}{J.~Mod.~Optics} \textbf{\bibinfo{volume}{41}},
  \bibinfo{pages}{2315} (\bibinfo{year}{1994}).

\bibitem[{\citenamefont{Bennett{ \it et~al.}}(1996)}]{bennett96b}
\bibinfo{author}{\bibfnamefont{C.~H.} \bibnamefont{Bennett{ \it et~al.}}},
  \bibinfo{journal}{Phys.~Rev.~Lett.} \textbf{\bibinfo{volume}{76}},
  \bibinfo{pages}{722} (\bibinfo{year}{1996}).

\bibitem[{\citenamefont{Kent}(1998)}]{kent98}
\bibinfo{author}{\bibfnamefont{A.}~\bibnamefont{Kent}},
  \bibinfo{journal}{Phys.~Rev.~Lett.} \textbf{\bibinfo{volume}{81}},
  \bibinfo{pages}{2839} (\bibinfo{year}{1998}).

\bibitem[{\citenamefont{Verstraete and Verschelde}(2003)}]{vv03}
\bibinfo{author}{\bibfnamefont{F.}~\bibnamefont{Verstraete}} \bibnamefont{and}
  \bibinfo{author}{\bibfnamefont{H.}~\bibnamefont{Verschelde}},
  \bibinfo{journal}{Phys.~Rev.~Lett.} \textbf{\bibinfo{volume}{90}},
  \bibinfo{pages}{097901} (\bibinfo{year}{2003}).

\bibitem[{\citenamefont{Gisin and Massar}(1997)}]{gisin97}
\bibinfo{author}{\bibfnamefont{N.}~\bibnamefont{Gisin}} \bibnamefont{and}
  \bibinfo{author}{\bibfnamefont{S.}~\bibnamefont{Massar}},
  \bibinfo{journal}{Phys.~Rev.~Lett.} \textbf{\bibinfo{volume}{79}},
  \bibinfo{pages}{2153} (\bibinfo{year}{1997}).

\bibitem[{\citenamefont{Ishizaka and Hiroshima}(2001)}]{ih01}
\bibinfo{author}{\bibfnamefont{S.}~\bibnamefont{Ishizaka}} \bibnamefont{and}
  \bibinfo{author}{\bibfnamefont{T.}~\bibnamefont{Hiroshima}},
  \bibinfo{journal}{Phys.~Rev.~A} \textbf{\bibinfo{volume}{62}},
  \bibinfo{pages}{022310} (\bibinfo{year}{2001}).

\bibitem[{\citenamefont{Verstraete et~al.}(2001)\citenamefont{Verstraete,
  Audenaert, and DeMoor}}]{vad01}
\bibinfo{author}{\bibfnamefont{F.}~\bibnamefont{Verstraete}},
  \bibinfo{author}{\bibfnamefont{K.}~\bibnamefont{Audenaert}},
  \bibnamefont{and} \bibinfo{author}{\bibfnamefont{B.}~\bibnamefont{DeMoor}},
  \bibinfo{journal}{Phys. Rev. A} \textbf{\bibinfo{volume}{64}},
  \bibinfo{pages}{012316} (\bibinfo{year}{2001}).

\bibitem[{\citenamefont{Munro et~al.}(2001)\citenamefont{Munro, James, White,
  and Kwiat}}]{munro01}
\bibinfo{author}{\bibfnamefont{W.~J.} \bibnamefont{Munro}},
  \bibinfo{author}{\bibfnamefont{D.~F.~V.} \bibnamefont{James}},
  \bibinfo{author}{\bibfnamefont{A.~G.} \bibnamefont{White}}, \bibnamefont{and}
  \bibinfo{author}{\bibfnamefont{P.~G.} \bibnamefont{Kwiat}},
  \bibinfo{journal}{Phys.~Rev.~A} \textbf{\bibinfo{volume}{64}},
  \bibinfo{pages}{R030302} (\bibinfo{year}{2001}).

\bibitem[{\citenamefont{Wootters}(1998)}]{wooters98}
\bibinfo{author}{\bibfnamefont{W.~K.} \bibnamefont{Wootters}},
  \bibinfo{journal}{Phys.~Rev.~Lett.} \textbf{\bibinfo{volume}{80}},
  \bibinfo{pages}{2245} (\bibinfo{year}{1998}).

\bibitem[{\citenamefont{Coffman et~al.}(2000)\citenamefont{Coffman, Kundu, and
  Wootters}}]{coffman00}
\bibinfo{author}{\bibfnamefont{V.}~\bibnamefont{Coffman}},
  \bibinfo{author}{\bibfnamefont{J.}~\bibnamefont{Kundu}}, \bibnamefont{and}
  \bibinfo{author}{\bibfnamefont{W.~K.} \bibnamefont{Wootters}},
  \bibinfo{journal}{Phys.~Rev.~A} \textbf{\bibinfo{volume}{61}},
  \bibinfo{pages}{052306} (\bibinfo{year}{2000}).

\bibitem[{\citenamefont{Bose and Vedral}(2000)}]{bose00}
\bibinfo{author}{\bibfnamefont{S.}~\bibnamefont{Bose}} \bibnamefont{and}
  \bibinfo{author}{\bibfnamefont{V.}~\bibnamefont{Vedral}},
  \bibinfo{journal}{Phys.~Rev.~A} \textbf{\bibinfo{volume}{61}},
  \bibinfo{pages}{R040101} (\bibinfo{year}{2000}).

\bibitem[{\citenamefont{Peters{ \it et~al.}}(2004)}]{peters04}
\bibinfo{author}{\bibfnamefont{N.~A.} \bibnamefont{Peters{ \it et~al.}}},
  \bibinfo{journal}{Phys.~Rev.~Lett.} \textbf{\bibinfo{volume}{92}},
  \bibinfo{pages}{133601} (\bibinfo{year}{2004}).

\bibitem[{\citenamefont{Werner}(1989)}]{werner}
\bibinfo{author}{\bibfnamefont{R.~F.} \bibnamefont{Werner}},
  \bibinfo{journal}{Phys.~Rev.~A} \textbf{\bibinfo{volume}{40}},
  \bibinfo{pages}{4277} (\bibinfo{year}{1989}).

\bibitem[{foo({\natexlab{a}})}]{footnotewerner}
\emph{\bibinfo{title}{{\rm Note that for certain entanglement and mixedness
  parameterizations, the Werner states {\it are} the MEMS~\cite{wei03a}}}}.

\bibitem[{\citenamefont{Zhang et~al.}(2002)\citenamefont{Zhang, Huang, Li, and
  Guo}}]{zhang02}
\bibinfo{author}{\bibfnamefont{Y.~S.} \bibnamefont{Zhang}},
  \bibinfo{author}{\bibfnamefont{Y.~F.} \bibnamefont{Huang}},
  \bibinfo{author}{\bibfnamefont{C.~F.} \bibnamefont{Li}}, \bibnamefont{and}
  \bibinfo{author}{\bibfnamefont{G.~C.} \bibnamefont{Guo}},
  \bibinfo{journal}{Phys.~Rev.~A} \textbf{\bibinfo{volume}{66}},
  \bibinfo{pages}{062315} (\bibinfo{year}{2002}).

\bibitem[{\citenamefont{Barbieri et~al.}(2004)\citenamefont{Barbieri,
  DeMartini, DiNepi, and Mataloni}}]{barbieri04}
\bibinfo{author}{\bibfnamefont{M.}~\bibnamefont{Barbieri}},
  \bibinfo{author}{\bibfnamefont{F.}~\bibnamefont{DeMartini}},
  \bibinfo{author}{\bibfnamefont{G.}~\bibnamefont{DiNepi}}, \bibnamefont{and}
  \bibinfo{author}{\bibfnamefont{P.}~\bibnamefont{Mataloni}},
  \bibinfo{journal}{Phys. Rev. Lett.} \textbf{\bibinfo{volume}{92}},
  \bibinfo{pages}{177901} (\bibinfo{year}{2004}).

\bibitem[{\citenamefont{Nielsen and Chuang}(2000)}]{nc00}
\bibinfo{author}{\bibfnamefont{M.~A.} \bibnamefont{Nielsen}} \bibnamefont{and}
  \bibinfo{author}{\bibfnamefont{I.~L.} \bibnamefont{Chuang}},
  \emph{\bibinfo{title}{Quantum Computationand Information}}
  (\bibinfo{publisher}{Cambridge University Press},
  \bibinfo{address}{Cambridge, U.~K.}, \bibinfo{year}{2000}).

\bibitem[{\citenamefont{Peters et~al.}(2004)\citenamefont{Peters, Wei, and
  Kwiat}}]{pwk04}
\bibinfo{author}{\bibfnamefont{N.~A.} \bibnamefont{Peters}},
  \bibinfo{author}{\bibfnamefont{T.-C.} \bibnamefont{Wei}}, \bibnamefont{and}
  \bibinfo{author}{\bibfnamefont{P.~G.} \bibnamefont{Kwiat}}, in
  \emph{\bibinfo{booktitle}{Proc. of SPIE Fluc. and Noise}}
  (\bibinfo{year}{2004}), vol. \bibinfo{volume}{5468}, pp.
  \bibinfo{pages}{269--281}.

\bibitem[{\citenamefont{Verstraete and Verschelde}(2002)}]{fv02}
\bibinfo{author}{\bibfnamefont{F.}~\bibnamefont{Verstraete}} \bibnamefont{and}
  \bibinfo{author}{\bibfnamefont{H.}~\bibnamefont{Verschelde}},
  \bibinfo{journal}{Phys. Rev. A} \textbf{\bibinfo{volume}{66}},
  \bibinfo{pages}{022307} (\bibinfo{year}{2002}).

\bibitem[{\citenamefont{Horodecki et~al.}(1997)\citenamefont{Horodecki,
  Horodecki, and Horodecki}}]{hhh97}
\bibinfo{author}{\bibfnamefont{M.}~\bibnamefont{Horodecki}},
  \bibinfo{author}{\bibfnamefont{P.}~\bibnamefont{Horodecki}},
  \bibnamefont{and}
  \bibinfo{author}{\bibfnamefont{R.}~\bibnamefont{Horodecki}},
  \bibinfo{journal}{Phys. Rev. Lett.} \textbf{\bibinfo{volume}{78}},
  \bibinfo{pages}{574} (\bibinfo{year}{1997}).

\bibitem[{foo({\natexlab{b}})}]{footnotesimulation}
\emph{\bibinfo{title}{{\rm In more detail, we project the ideal target state
  into 16 basis vectors, such as $\langle 00|$, $\langle11|$, $\langle
  (0+i1)0|$, etc., to obtain a list of probabilities of given ``measurement''
  outcomes. These probabilities are then multiplied by a constant number
  simulating an expected average number of counts in a total basis measurement,
  e.g., what one would expect to observe when projecting into $\langle 00|$,
  $\langle 01|$, $\langle 10|$, and $\langle 11|$. Next, each of these ideal
  counts (plus one to avoid zero distributions) is used as the mean of a
  Poisson distribution, from which a random number is generated. These
  ``measurement'' values are then processed using a maximum likelihood
  technique to give a physically valid perturbed density
  matrix~\cite{james01,ajk04}. If the fidelity between the perturbed density
  matrix and the target state is greater than 0.9900, the tangle and linear
  entropy are calculated and plotted in Fig.~\ref{memsFcalc}.}}}

\bibitem[{\citenamefont{Peters{ \it et~al.}}(2003)}]{peters03}
\bibinfo{author}{\bibfnamefont{N.}~\bibnamefont{Peters{ \it et~al.}}},
  \bibinfo{journal}{J. Quant. Inf. and Comp.} \textbf{\bibinfo{volume}{3}},
  \bibinfo{pages}{503} (\bibinfo{year}{2003}).

\bibitem[{\citenamefont{Knill et~al.}(1998)\citenamefont{Knill, Laflamme, and
  Zurek}}]{klz98}
\bibinfo{author}{\bibfnamefont{E.}~\bibnamefont{Knill}},
  \bibinfo{author}{\bibfnamefont{R.}~\bibnamefont{Laflamme}}, \bibnamefont{and}
  \bibinfo{author}{\bibfnamefont{W.~H.} \bibnamefont{Zurek}},
  \bibinfo{journal}{Science} \textbf{\bibinfo{volume}{279}},
  \bibinfo{pages}{574} (\bibinfo{year}{1998}).

\end{thebibliography}

\end{document}